\documentclass[10pt,aps,prl,twocolumn,superscriptaddress]{revtex4}
\usepackage[linktocpage=true, colorlinks=true, urlcolor=blue, linkcolor=blue, citecolor=blue]{hyperref}
\usepackage{graphicx}
\usepackage{amsmath,amssymb}
\usepackage{color}
\usepackage{subfigure}
\usepackage{url}

\graphicspath{{../Images/}}

\begin{document}

\title{Real-time digital holography of the retina by principal component analysis}

\author{Leo Puyo}
\author{Loic Bellonnet-Mottet}
\author{Antoine Martin}
\author{Francois Te}
\author{Michel Paques}
\affiliation{
Centre Hospitalier National d’Ophtalmologie des Quinze-Vingts, INSERM CIC 1423. 28 rue de Charenton, 75012 Paris France
}
\author{Michael Atlan}
\affiliation{
Centre National de la Recherche Scientifique (CNRS) UMR 7587, Institut Langevin. Paris Sciences et Lettres (PSL) University, Universit\'e Pierre et Marie Curie (UPMC), Universit\'e Paris 7. \'Ecole Sup\'erieure de Physique et de Chimie Industrielles ESPCI Paris - 1 rue Jussieu. 75005 Paris. France
}

\date{\today}

\begin{abstract}
We demonstrate the feasibility of high-quality digital holography of the human retina in real-time with a fast camera and commodity computer hardware. High throughput rendering of digital Fresnel holograms from optically-acquired inline interferograms is performed in conjunction with temporal demodulation by projection of hologram sequences onto a data-derived basis in order to discriminate local narrowband coherent detection contrasts, mostly due to blood flow and optical absorption, from spurious interferometric contributions. Digital holograms are calculated from a sustained input stream of 16-bit, 1024-by-1024-pixel interferograms recorded at up to 500 frames per second, processed by principal component analysis. This temporal signal demodulation scheme consists in the projection of stacks of 32 consecutive holograms onto a basis calculated by eigendecomposition of the matrix of their time-lagged covariance; it is performed up to 20 times per second with commodity computer hardware.
\end{abstract}

\maketitle

Fast, robust, efficient and versatile digital image acquisition and real-time rendering software with sustained high data throughput is a key requirement for the development of digital holographic imaging of the eye fundus in clinical settings. The technical issues for real-time digital holography of the retina are manifold. Interferometric detection must be performed in low-light, and hence in high coherent gain regime, in order to comply with the safety limits of eye exposure to near infrared radiation. Additionally, inline interferometry may be chosen over an off-axis configuration to make use of the full spatial bandwidth of the sensor array for imaging, which comes at a price : self-beating and twin-image spurious contributions strongly decrease the quality of the reconstructed image.\\

The available ultrahigh-speed cameras with sensor arrays enabling interferogram sampling at tens of thousands of frames per second cannot be used for real-time imaging because of the lack of suitable data interface. This jeopardizes the calculation of high quality Doppler-contrast movies via short-time Fourier transforms to reveal local temporal fluctuations~\cite{Puyo2018}. At lower frame rates, time-averaged digital holography reveals high frequency Doppler broadening beyond Shannon-Nyquist sampling limit; the observed narrowband coherent detection contrasts are mostly due to blood flow and optical absorption~\cite{Puyo2020low}. However, cameras enabling interferogram sampling at hundreds of frames per second with compatible frame grabbing interfaces for real-time data transfer have become available. Yet their frame rate is still insufficient to perform retinal holography by Fourier transform temporal demodulation due to the presence of a large footprint of spurious signals.\\

Temporal demodulation consists in a linear combination of a sequence of recorded interferograms, whose coefficients are the ones of a discrete Fourier transform~\cite{Freischlad1990}. When the modulation is not known, principal component analysis of interferograms~\cite{Vargas2011} may be used for non-parametric, non-iterative demodulation. It identifies uncorrelated variables in the structure of the data, the principal components, from the analysis of its correlations. It is the spatial average counterpart of the decomposition of local digital hologram variations with time demonstrated for low-light temporal demodulation in the presence of random phase drifts~\cite{Lopes2015}. A window of opportunity for retinal imaging by inline digital holography in real-time is opened with such data-driven adaptive temporal signal demodulation on a graphics processing unit (GPU). Eigendecomposition of the matrix of time-lagged covariance of consecutive holograms allows to construct a demodulation basis suitable for cancellation of spurious interferometric contributions, uncorrelated to signal variations. This solution can be used for temporal fluctuation analysis to improve significantly image quality in retinal imaging.\\

\begin{figure}[]
\centering
\includegraphics[width = 8.0 cm]{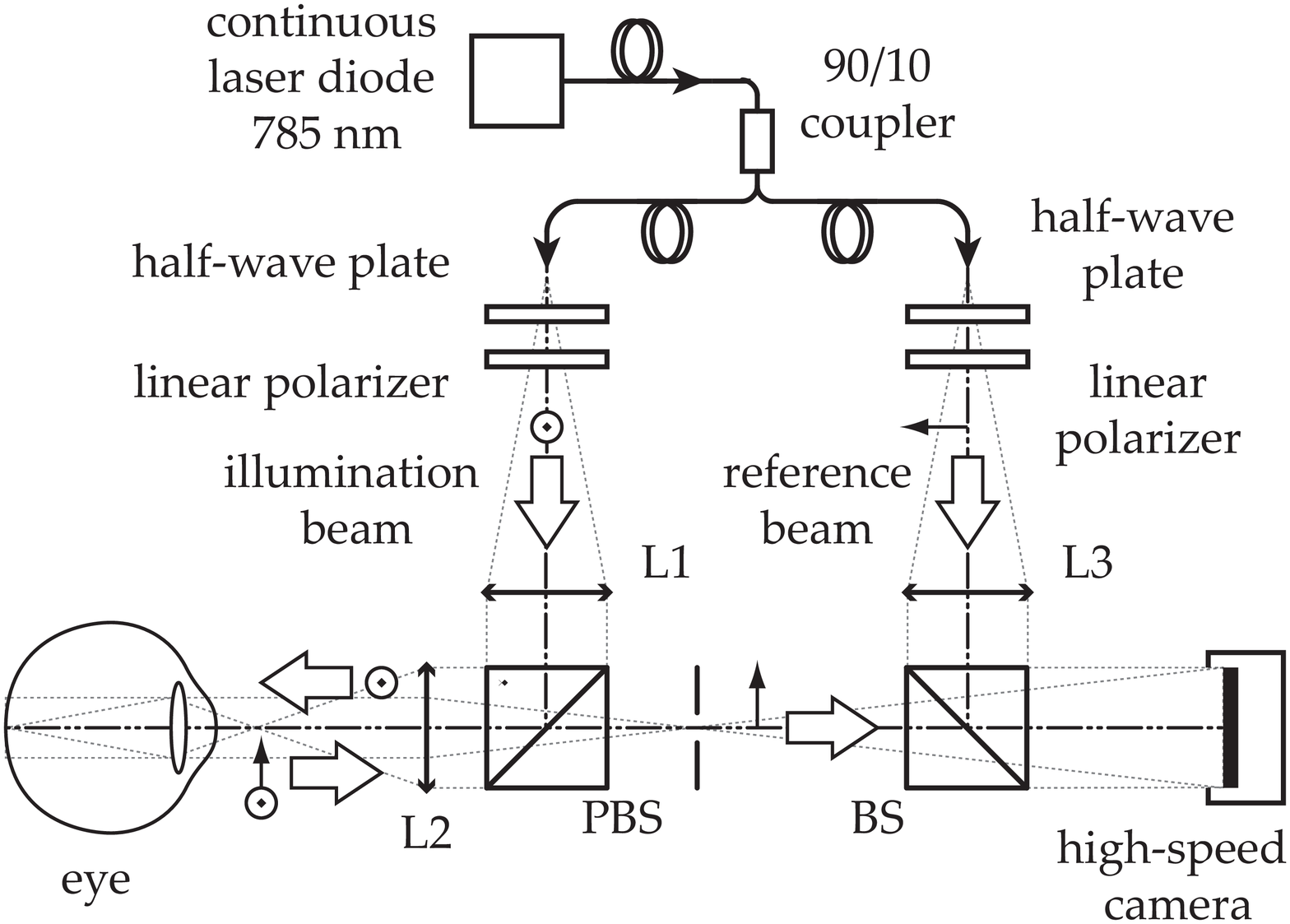}
\caption{Sketch of the laser inline Mach-Zenhder interferometer used to record digital interferograms of the light backscattered by the eye fundus of a volunteer.}
\label{fig_Setup}
\end{figure}

The experimental Mach-Zehnder inline optical interferometer used for this study is sketched in Fig.~\ref{fig_Setup}. The laser source used for the experiments was a $\sim$ 50 mW, single-mode, fibered diode at wavelength $\lambda = 785$ nm, and optical frequency $\omega_{\rm L} / (2 \pi) = 3.8 \times 10^{14} \, \rm Hz$ (Thorlabs LP785-SAV50 785 nm, 50 mW, VHG Wavelength-Stabilized SF Laser Diode, Internal Isolator). The retina of one healthy volunteer was illuminated with a continuous-wave laser beam of $\sim$ 1.3 mW power, focused at the natural focus point of the eye. This irradiation level is compliant with the International Organization for Standardization norm ISO 15004-2:2007. Informed consent was obtained from the subject, experimental procedures adhered to
the tenets of the Declaration of Helsinki, study authorization
was obtained from the appropriate local ethics review boards
CPP and ANSM, and the clinical trial was registered under the references IDRCB 2019-A00942-5, and NCT04129021.\\

In the object arm, a polarizing beam splitter cube was used to illuminate the eye with linearly polarized light and collect the cross-polarized backscattered component, in order to avoid specular reflections. The backscattered optical field $E$ was mixed with a separate reference (local oscillator) field $E_{\rm LO}$ with a non-polarizing beam splitter cube to ensure inline interferometry. Light-tissue interaction resulted in a local phase variation $\phi(t)$ of the backscattered laser optical field $E(t) = {\cal E}(t)  \exp \left[ i \omega_{\rm L} t + i \phi(t) \right]$, which was mixed with the LO field from the reference channel $E_{\rm LO}(t) = {\cal E}_{\rm LO}(t) \exp \left[ i \omega_{\rm L}t + i \phi_{\rm LO}(t)\right]$. The quantity $i$ is the imaginary unit, ${\cal E}_{\rm LO}(t)$, ${\cal E}(t)$ are the envelopes, and $\exp[i\phi_{\rm LO}(t)]$, $\exp[i\phi(t)]$ are the phase factors of the fields, respectively. In this description, we consider a reference wave devoid of temporal fluctuations in amplitude and phase in the (filtered) detection bandwidth, so we can write $E_{\rm LO}(t) = {\cal E}_{\rm LO} \exp \left[ i \omega_{\rm L}t \right]$, where ${\cal E}_{\rm LO}$ is constant in time; in practice, its amplitude fluctuations create spurious low-frequency interferometric contribution within the camera bandwidth. Optical interferograms of $1024 \times 1024$ pixels of coordinates $(x,y)$ were digitally acquired by an Adimec Quartz Q-2A750-Hm/CXP-6 camera, with a Bitflow Cyton-CXP framegrabber, at a frame rate of $\omega_{\rm S} / (2 \pi) = 500 \, \rm Hz$, with a pixel size $d = 12 \, \mu \rm m$. The distance between the eye and the sensor was $\sim 40 \, \rm cm$.\\

\begin{figure}[]
\centering
\includegraphics[width = 8.0 cm]{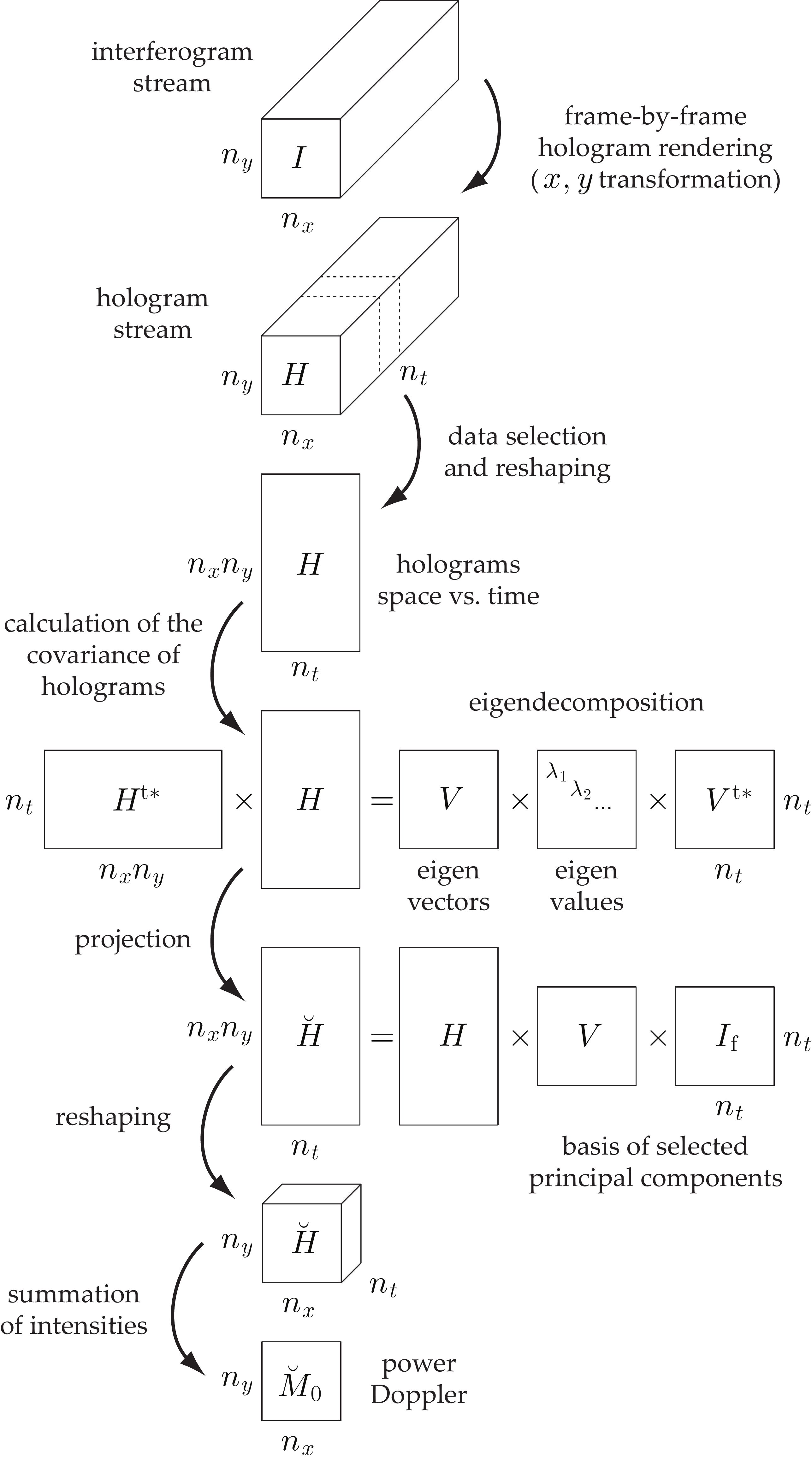}
\caption{Flowchart of the data processing steps for temporal demodulation of retinal holograms by eigendecomposition.
}
\label{fig_EigendecompositionFlowchart}
\end{figure}

The Fresnel transformation was used for image rendering because it requires only one spatial Fourier transform per reconstructed hologram
\begin{eqnarray}\label{eq_FresnelTransform}
\nonumber H(x,y,t) = \frac{i}{\lambda z}\exp \left( -i k z \right) \iint I(x',y',t)\\
\times \exp \left[\frac{-i \pi}{\lambda z} \left((x-x')^2 + (y-y')^2\right) \right] {\rm d}x' {\rm d}y'
\end{eqnarray}
where the parameter $z$ corresponds to the sensor-to-object distance for a flat reference wavefront and in the absence of lens in the object path. Because of the inline interferometric configuration, the cross-beating component $H = {\cal E}_{\rm LO}^* {\cal E} \exp \left( i \phi \right)$ of the interferogram $I = \left| {\cal E} \right|^2 + \left| {\cal E}_{\rm LO} \right|^2 + H + H^*$ is not separated from the self-beating components $\left| {\cal E} \right|^2 + \left| {\cal E}_{\rm LO} \right|^2$ and the twin cross-beating component $H^*$ in the Fourier reciprocal space. The asterisk $^*$ denotes the complex conjugate.\\

In our previous work, imaging of blood flow was done by calculating the temporal Fourier transform of the holograms over a short-time window of a few hundreds holograms; power Doppler images were calculated by high-pass filtering of the Fourier spectrum, and integration of its envelope~\cite{Puyo2018}. Yet with a slower camera, this procedure is not sufficient to reveal high-quality retinal images in real-time. An additional filtering operation by singular value decomposition, adapted from ultrasound data processing~\cite{Demene2015}, was introduced prior to the calculation of the temporal Fourier transform~\cite{Puyo2020svd} to cancel spurious signal contributions : For a given short-time window of $n_t$ consecutive holograms of spatial dimension $(n_x, n_y)$, the stack of holograms is reshaped into a bidimensional space-time matrix $H$ of size ($n_x n_y$, $n_t$), where each column of the matrix contains all the hologram pixels at a given time. It admits a singular value decomposition such that
\begin{equation} \label{eq:eq_SingularValueDecompositionH}
H = U \, \Delta \, V^{\rm t *}
\end{equation}
where $^{t*}$ stands for the conjugate transpose. The rectangular and diagonal matrix $\Delta$ of size ($n_x n_y$, $n_t$) contains $n_t$ real positive singular values sorted by increasing order of magnitude, and $U$ and $V$ are unitary matrices ($U\,U^{t*} = I$, and $V\,V^{t*} = I$, where $I$ is an identity matrix) whose columns are called spatial and temporal singular vectors, of sizes ($n_x n_y$, $n_x n_y$) and ($n_t$, $n_t$), respectively. Singular value filtering of $H$ consists of forming $H_{\rm f}$ such that
\begin{equation} \label{eq:eq_SingularValueDecompositionHf}
H_{\rm f} = U \, \Delta \, I_{\rm f} \, V^{\rm t *}
\end{equation}
where the diagonal matrix $I_{\rm f}$, whose values are 1 between the indexes $n_1$ and $n_2$ and 0 elsewhere is used to cancel the highest singular values in $\Delta$. Although the results are valuable because the detectivity of weak local Doppler signals is improved~\cite{Puyo2020svd}, this filter (and even more so its combination with temporal Fourier transformation) is too compute-intensive for real-time imaging with commodity hardware.\\

To circumvent this issue, another projection basis whose calculation only requires the eigendecomposition of a matrix of size ($n_t, n_t$) is calculated. The methodology is sketched in Fig.~\ref{fig_EigendecompositionFlowchart}. We first form the square, complex-valued and symmetric time-lagged covariance matrix of consecutive holograms
\begin{equation} \label{eq:eq_C}
C = H^{t*}\, H 
\end{equation}
The matrix $C$ of size $(n_t, n_t)$ can be diagonalized using the unitary matrix $V$ such that the eigendecomposition of $C$ is a product of 3 matrices $V$, $\Lambda$, and $V^{\rm t *}$ (Takagi's factorization). By expressing $H$ with Eq.~\ref{eq:eq_SingularValueDecompositionH}, Eq.~\ref{eq:eq_C} takes the form
\begin{equation} \label{eq:eq_EigendecompositionC}
C = V \, \Lambda \, V^{\rm t *}
\end{equation}
The square diagonal matrix $\Lambda = \Delta^2$ of size ($n_t, n_t$) contains $n_t$ real positive eigenvalues, sorted by increasing value, and associated to the temporal singular vectors embedded in the columns of the matrix $V$.\\

The projection of $H$ onto the matrix of temporal singular vectors $V$ forms its uncorrelated principal components~\cite{Vargas2011}
\begin{equation} \label{eq:eq_PCA}
\breve{H} = H \, V \ I_{\rm f}
\end{equation}
The columns of $\breve{H}$ are the principal components of $H$, selected by the diagonal matrix $I_{\rm f}$. It is reshaped into a matrix $\breve{H}(x,y,\varpi)$ of size $(n_x,n_y,n_t)$ whose envelope is summed over $\varpi$ to create a power Doppler image $\breve{M}_{0}(x,y)$ of the retina
\begin{equation}\label{eq:eq_PowerDopplerBreve}
\breve{M}_{0}(x,y) = \sum_\varpi \left|\breve{H}(x,y,\varpi) \right|^2
\end{equation}
Note that the variables $t$ and $\varpi$ are conjugate variables for reshaped matrices $H(x,y,t)$ and $\breve{H}(x,y,\varpi)$ (Eq.~\ref{eq:eq_PCA}), in the same manner that $t$ and $\omega$ are conjugate variables for $H(x,y,t)$ and $\tilde{H}(x,y,\omega)$ (Eq.~\ref{eq:eq_FFT}).\\

\begin{figure}[]
\centering
\includegraphics[width = 8cm]{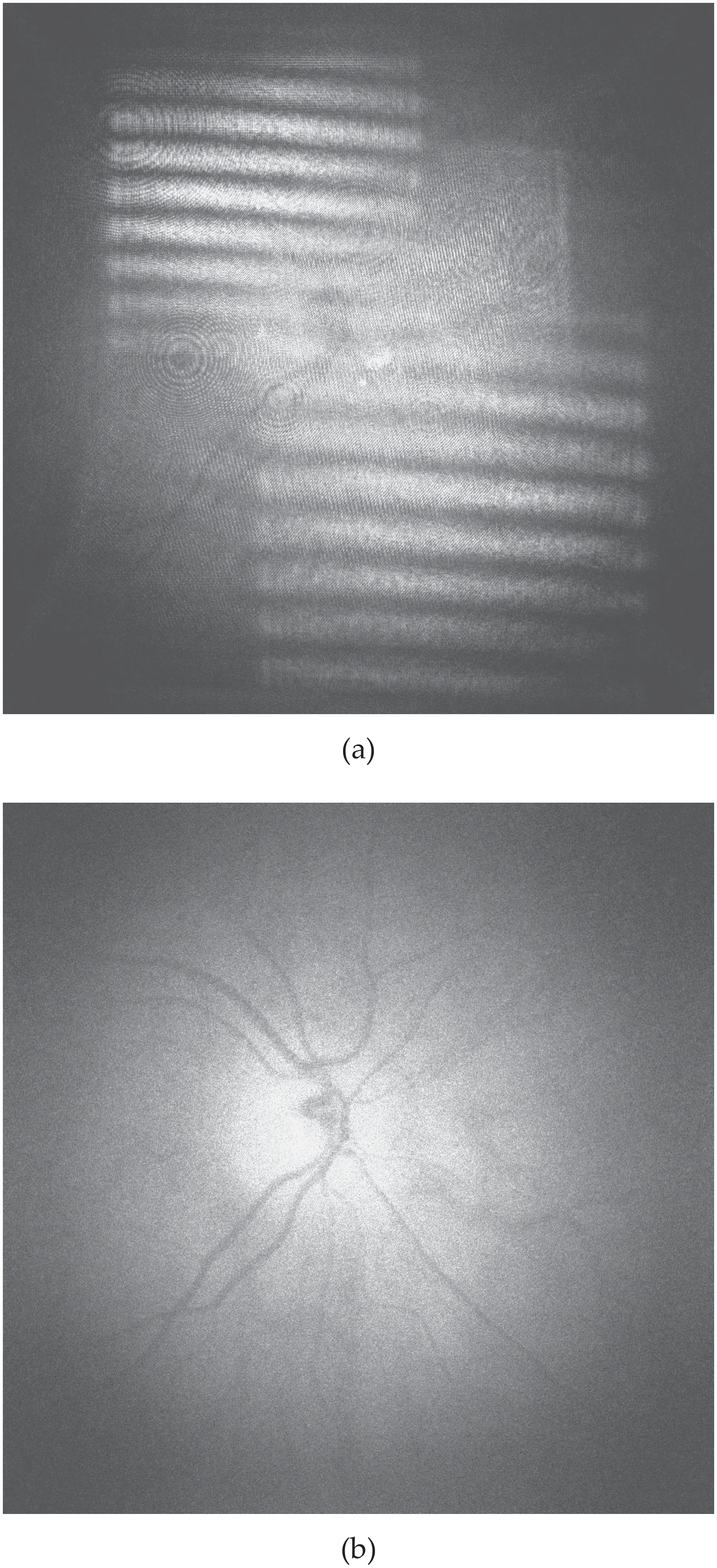}
\caption{Real-time visualization of inline digital holograms of a human retina from an input stream of 16-bit, 1024-by-1024-pixel interferograms at a rate of 500 frames per second. The temporal signal demodulation of 32 consecutive holograms stacks is done by projection onto a Fourier basis (a) and onto a basis derived from eigendecomposition of the matrix of time-lagged covariance of consecutive holograms (b), at a rate of 20 Hz. \href{https://youtu.be/0uwCdAddRns}{Vizualization 1}.}
\label{fig_HolovibesRetinaSVD}
\end{figure}

Eq.~\ref{eq:eq_PCA} is the counterpart of a temporal demodulation by discrete Fourier transformation followed by a frequency filter, which can be written
\begin{equation} \label{eq:eq_FFT}
\tilde{H} = H \, W \, I_{\rm f}
\end{equation}
where $W = (w_{mn})$ is a matrix of size $(n_t,n_t)$, with $w_{mn} = \exp{\left[-2i(m-1)(n-1)\pi/n_t\right]}/\sqrt{n_t}$ ($m\in\{1,...,n_t\}, n\in\{1,...,n_t\}$). The diagonal matrix $I_{\rm f}$, whose diagonal values are 1 between the indexes $n_1$ and $n_2$ and 0 elsewhere, is used as a high-pass filter to cancel low temporal frequency contributions. The envelope of $\tilde{H}$ is summed over temporal frequencies $\omega$ to create a power Doppler image
\begin{equation}\label{eq:eq_PowerDoppler}
M_{0}(x,y) = \sum_\omega \left|\tilde{H}(x,y,\omega) \right|^2
\end{equation}
The empirical choice of the index span of diagonal values set to 1 in $I_{\rm f}$ defines a high-pass filter used to maximize $M_{0}$ image quality. In practice, setting $n_1 = 8$ and $n_2 = 24$ in a temporal demodulation by discrete Fourier transformation (Eq.~\ref{eq:eq_FFT}) of $n_t = 32$ consecutive holograms gives the result displayed in Fig.~\ref{fig_HolovibesRetinaSVD}(a), and in \href{https://youtu.be/0uwCdAddRns}{Vizualization 1}. In the same way, the choice of threshold indexes $n_1$ and $n_2$ in $I_{\rm f}$ to truncate the eigenvectors matrix $V$ (Eq.~\ref{eq:eq_PCA}) is made empirically, by maximization of image quality. The index $n_1$ may always be set to 1 in order to keep the eigenvalues of lowest order of magnitude; but if $n_2$ is set too high, the clutter may not be rejected. In practice, setting $n_1 = 1$ and $n_2 = 20$ in a eigendecomposition of $n_t = 32$ consecutive holograms gives the result displayed in Fig.~\ref{fig_HolovibesRetinaSVD}(b), and in \href{https://youtu.be/0uwCdAddRns}{Vizualization 1}. It is worth remarking that image normalization by their respective average value is crucial to cancel the strong $M_0$ and $\breve{M}_{0}$ signal variations during visualization. For comparison, the same interferogram footage $I$ was used to calculate $M_0$ and $\breve{M}_{0}$ in Fig.~\ref{fig_HolovibesRetinaSVD}. Much better spurious signal filtering is performed with eigendecomposition than with Fourier demodulation. Throughput benchmarks for both methods are reported in Table~\ref{table_Benchmarks}.\\

%


Real-time hologram rendering with temporal demodulation by principal component analysis is implemented in the software holovibes 8.0, written in C++ and compiled with CMake 3.5 and NVIDIA CUDA toolkit 10.1. All calculations are performed by a NVIDIA Titan RTX GPU card on single precision floating-point complex-valued arrays of type \emph{cuComplex}. The acquisition of the raw interferogram stream from the camera is bufferized to avoid any frame drop, in order to ensure the consistency of the temporal demodulation, either via Eq.~\ref{eq:eq_FFT} or Eq.~\ref{eq:eq_PCA}. The propagation integral (Eq.~\ref{eq_FresnelTransform}) and the Fourier transform (Eq.~\ref{eq:eq_FFT}) are computed by the functions from the \emph{cuFFT} library. The \emph{cuBLAS} and \emph{cuSOLVER} libraries were used for matrix operations. In particular, the construction (Eq.~\ref{eq:eq_C}) and eigendecomposition (Eq.~\ref{eq:eq_EigendecompositionC}) of the covariance matrix  are performed by the functions \emph{cublasCgemm()} and \emph{cusolverDnCheevd()}, respectively.\\

\begin{table}[]
\centering
\begin{tabular}{|c|c|c|}
\hline
\, demodulation \, & \, input throughput \, & \, output throughput \,  \\
\hline
\hline
\, FFT \, & \, 1.0 GB/s \, & \, $\sim$17 Gvoxel/s \, \\
\hline
\, EIG \, & \, 1.0 GB/s \, & \, $\sim$671 Mvoxel/s \, \\
\hline
\end{tabular}
\caption{Sustained input/output throughput with no frame dump, benchmarked with a Titan RTX GPU. Input data throughput : 16-bit, $1024 \times 1024$-pixel interferograms at 500 frames per second. Hologram rendering : Fresnel transformation (Eq.~\ref{eq_FresnelTransform}). Temporal demodulation by short-time Fourier transform (FFT) on 32 consecutive holograms (Eq.~\ref{eq:eq_FFT}), versus temporal demodulation by projection of stacks of 32 consecutive holograms onto a data-derived basis (Eq.~\ref{eq:eq_PCA}) calculated by eigendecomposition (EIG) of the matrix of time-lagged covariance of holograms. Maximum output data throughput : $32 \times 1024 \times 1024$-voxel holograms $\tilde{H}$ at 500 frames per second for FFT, and $32 \times 1024 \times 1024$-voxel holograms $\breve{H}$ at 20 frames per second for EIG.}\label{table_Benchmarks}
\end{table}
In conclusion, we have demonstrated real-time computation and visualization of high-quality inline digital holograms of the eye fundus from an input stream of 16-bit, 1024-by-1024-pixel interferograms recorded at 500 frames per second, by Fresnel transformation. The stream of holograms is processed by principal component analysis of stacks of 32 consecutive holograms. This approach enables to perform clutter-free digital holographic imaging of the retina with at a rate of 20 frames per second on commodity computer hardware.\\

This work was supported by the IHU FOReSIGHT (ANR-18-IAHU-01), the European Research Council (ERC Synergy HELMHOLTZ, agreement \#610110), and by the Sesame program of the Region Ile-de-France (4DEye).

\bibliographystyle{unsrt}

\begin{thebibliography}{10}

\bibitem{Puyo2018}
L.~Puyo, M.~Paques, M.~Fink, J.-A. Sahel, and M.~Atlan.
\newblock In vivo laser doppler holography of the human retina.
\newblock {\em Biomedical Optics Express}, 9(9):4113--4129, Sep 2018.

\bibitem{Puyo2020low}
Leo Puyo, Michel Paques, and Michael Atlan.
\newblock Low frame rate laser doppler holography, 2020.

\bibitem{Freischlad1990}
K.~Freischlad and C.L. Koliopoulos.
\newblock Fourier description of digital phase-measuring interferometry.
\newblock {\em Journal of the Optical Society of America A: Optics and Image
  Science, and Vision}, 7(4):542--551, 1990.

\bibitem{Vargas2011}
Javier Vargas, J~Antonio Quiroga, and T~Belenguer.
\newblock Phase-shifting interferometry based on principal component analysis.
\newblock {\em Optics letters}, 36(8):1326--1328, 2011.

\bibitem{Lopes2015}
Fernando Lopes and Michael Atlan.
\newblock Singular-value demodulation of phase-shifted holograms.
\newblock {\em Optics letters}, 40(11):2541--2544, 2015.

\bibitem{Demene2015}
Charlie Demen{\'e}, Thomas Deffieux, Mathieu Pernot, Bruno-F{\'e}lix Osmanski,
  Val{\'e}rie Biran, Jean-Luc Gennisson, Lim-Anna Sieu, Antoine Bergel,
  Stephanie Franqui, Jean-Michel Correas, et~al.
\newblock Spatiotemporal clutter filtering of ultrafast ultrasound data highly
  increases doppler and fultrasound sensitivity.
\newblock {\em IEEE transactions on medical imaging}, 34(11):2271--2285, 2015.

\bibitem{Puyo2020svd}
Leo Puyo, Michel Paques, and Michael Atlan.
\newblock Spatio-temporal filtering in laser doppler holography, 2020.

\end{thebibliography}

\end{document}